\documentclass[10pt, conference]{IEEEtran}

\usepackage{url}
\usepackage{cite}
\usepackage{xspace}
\usepackage{hyperref}
\usepackage{makecell}
\usepackage{graphicx}
\usepackage{algorithm}
\usepackage{algpseudocode}
\usepackage{color, xcolor}
\usepackage{amsmath, amssymb, amsfonts}

\newcommand{\etc}{\textit{etc.}\xspace}
\newcommand{\etal}{\textit{et al.}\xspace}

\newcommand{\tabref}[1]{TABLE~\ref{#1}\xspace}
\newcommand{\figref}[1]{Fig.~\ref{#1}\xspace}
\newcommand{\secref}[1]{Section~\ref{#1}\xspace}
\newcommand{\algopref}[1]{Algorithm~\ref{#1}\xspace}

\newcommand{\deepft}{{\sc DeepFeature}\xspace}
\newcommand{\approach}{{\sc DeepPrior}\xspace}

\newcommand{\similarity}{{\sc DeepSimilarity}\xspace}
\newcommand{\bft}{\textbf{\textit{Bug Feature}}\xspace}
\newcommand{\approachExp}{{\sc \textbf{DeepPrior}}\xspace}
\newcommand{\cft}{\textbf{\textit{Context Feature}}\xspace}

\newcommand{\tabincell}[2]{\begin{tabular}{@{}#1@{}}#2\end{tabular}}

\begin{document}

\title{Prioritize Crowdsourced Test Reports \\ via Deep Screenshot Understanding}

\author{Anonymous Author(s)}

 \author{\IEEEauthorblockN{Shengcheng Yu, Chunrong Fang$^*$, Zhenfei Cao, Xu Wang, Tongyu Li, Zhenyu Chen}
 	\IEEEauthorblockA{State Key Laboratory for Novel Software Technology, Nanjing University, China \\
 	$^*$Corresponding author: fangchunrong@nju.edu.cn}}

\maketitle

\begin{abstract}

Crowdsourced testing is increasingly dominant in mobile application (app) testing, but it is a great burden for app developers to inspect the incredible number of test reports. Many researches have been proposed to deal with test reports based only on texts or additionally simple image features. However, in mobile app testing, texts contained in test reports are condensed and the information is inadequate. Many screenshots are included as complements that contain much richer information beyond texts. This trend motivates us to prioritize crowdsourced test reports based on a deep screenshot understanding. 

In this paper, we present a novel crowdsourced test report prioritization approach, namely \approach. We first represent the crowdsourced test reports with a novelly introduced feature, namely \deepft, that includes all the widgets along with their texts, coordinates, types, and even intents based on the deep analysis of the app screenshots, and the textual descriptions in the crowdsourced test reports. \deepft includes the \bft, which directly describes the bugs, and the \cft, which depicts the thorough context of the bug. The similarity of the \deepft is used to represent the test reports' similarity and prioritize the crowdsourced test reports. We formally define the similarity as \similarity. We also conduct an empirical experiment to evaluate the effectiveness of the proposed technique with a large dataset group. The results show that \approach is promising, and it outperforms the state-of-the-art approach with less than half the overhead.

\end{abstract}


\section{Introduction}
\label{sec:intro}

Crowdsourcing has been one of the mainstream techniques in many areas. The openness of crowdsourcing brings many advantages. For example, the operations on crowdsourcing subjects can be simulated in multiple different real practical environments. Such advantages help alleviate the severe ``fragmentation problem'' in mobile application (app) testing \cite{yu2019lirat}. There are hundreds of thousands of different mobile device models with different brands, operating system (OS) versions, and hardware sensors, which is the well-known ``fragmentation problem'' in Android testing. Crowdsourced testing is one of the best solutions faced with such a problem. App developers can distribute their apps to crowdworkers with different mobile devices and require them to submit test reports containing \textbf{app screenshots} and \textbf{textual descriptions}. This helps app developers reveal as many problems as possible.

However, report reviewing efficiency in crowdsourced testing is a severe problem. The openness of crowdsourcing can lead to a great number of reports being submitted, and almost 82\% of the submitted reports are duplicate \cite{wang2019images}. It is quite tough work to review all the reports automatically due to the complexity. In the text part, the complexity of neural language can lead to ambiguity, and crowdworkers may use different words to describe the same objects or use one word to describe completely different scenarios. In the image part, screenshot similarity can also provide little help because many app functions share similar UI. Therefore, it is hard but important for app developers to reveal all the mentioned bugs as early as possible.

Among the recent researches, test reports' disposal is always divided into two parts: app screenshots and textual descriptions. Existing researches analyze these two parts separately to extract features. For textual descriptions, existing approaches extract the keywords and normalize the keywords according to predefined vocabulary. For app screenshots, they treat each screenshot as a whole and extract the image features represented with numeric vectors. After obtaining the results from two parts, most researches currently rely on texts and consider screenshots as supplemental materials or simply concatenate image information and text information. However, we think this kind of disposal can cause much valuable information to be missing. The relationship between textual descriptions and app screenshots is left out, and the report deduplication or prioritization can be less effective.

In this paper, we propose a novel approach, namely \approach, to {\sc Prior}itize crowdsourced test reports via {\sc Deep} screenshot understanding. \approach considers the deep understanding of both app screenshots and textual descriptions in detail. For a submitted test report, we extract information from both screenshots and texts. In screenshots, we collect all the widgets according to computer vision (CV) technologies, and we locate the problem widget (denoted as $W_P$) according to the textual descriptions (details in \secref{sec:image}). The remaining widgets are treated as context widgets (denoted as $W_C$). Texts are processed with Natural Language Processing (NLP) technologies and are divided into two parts: the reproduction steps (denoted as $R$) and the bug description (denoted as $P$). The reproduction steps are further normalized into ``action-object'' sequences. The bug description is also further processed to extract the problem widget description for $W_P$ localization.

Instead of processing app screenshots and textual descriptions separately, we take them as a whole and collect all the information as a \deepft for a report. Based on the relativity to the bug itself, \deepft includes \bft (BFT) and \cft (CFT). The \bft consists of $W_P$ and $P$, and it represents the information directly relevant to the bug revealed in the report. The \cft consists of $W_C$ and $R$, and it represents the context information, including the operation track triggering the bug and the activity information where the bug occurs.

After integrating the above features into the \deepft, \approach calculates the \similarity among all the reports for prioritization. For \bft and \cft, we calculate \similarity separately.

For \bft, to calculate the \similarity of the $W_P$ in the reports, we use CV technologies to extract and match the feature points. $P$ is a short textual description, so we use NLP technologies to extract the bug-related keywords based on our self-built vocabulary, and compare the keyword frequency as the \similarity.

For \cft, $W_C$ is fed into a pre-trained deep learning classifier to identify each widget's type, and the number vector for each type is considered as the $W_C$ \similarity. $R$ is composed of a series of actions and the corresponding widgets, representing the sequence from the app launching to the bug occurrence. Therefore, we extract the actions and the objects using NLP technologies in the $R$ order. We take the ``action-object'' sequence as the operation track and calculate the \similarity. 

Then starts the prioritization. We first construct a \textit{NULL Report} (defined in \secref{sec:prioritization}) and append it to the prioritized report pool. Then, we repeatedly calculate the \similarity between each unprioritized report and all the reports in the prioritized report pool. The report with the lowest ``minimum \similarity'' with all the reports in the prioritized report pool is put into the prioritized report pool.

We also design an empirical experiment, using a large-scale dataset group from a large and active crowdsourced testing platform\footnote{anonymous for double-blind principle}. We compare \approach with two other strategies, and the results show that \approach is effective.

The noteworthy contributions of this paper are as follows.

\begin{itemize}
	\item We propose a novel approach that prioritizes crowdsourced test reports via deep screenshot understanding and detailed text analysis. We extract all the widgets from the screenshots, classify textual information to different categories and form the \deepft.
	\item We construct an integrated dataset group for deep screenshot understanding, including a large-scale widget image dataset, a large-scale test report keyword vocabulary, a large-scale text classification dataset, and a large-scale crowdsourced test report dataset.
	\item Based on the dataset group, we conduct an empirical evaluation of the proposed approach \approach, and the results show that \approach outperforms the state-of-the-art approach with less than half the overhead.
\end{itemize}

\textbf{More resources can be found on our online package}\footnote{\url{https://sites.google.com/view/deepprior}}.

\section{Background \& Motivation}
\label{sec:bg&mt}

Crowdsourced testing has gained a large amount of popularity in mobile app testing, its advantages are obvious, but its drawbacks are also unignorable. On most mainstream crowdsourced testing platforms, crowdworkers are required to submit a report to describe the bug they meet. The main body of a report is a screenshot of the bug and a textual description. The app screenshot and the textual description are also the principal basis to prioritize the crowdsourced test reports.

Current solutions for crowdsourced test report processing that consider the screenshots like \cite{wang2019images}\cite{feng2016multi} mainly analyze the app screenshot features and textual description information to measure the similarity among all the reports. Though they consider the app screenshots, they simply treat the images as $width \times height \times RGB$ matrixes. However, these approaches ignore the rich and valuable information, and we hold the opinion that the app screenshots should be viewed as a collection of meaningful widgets instead of the collection of meaningless pixels. We make such a stand because while reviewing the crowdsourced test report dataset, we find some vivid examples that the existing approaches have difficulty handling because they merely make simple feature extraction instead of deep screenshot understanding.

\subsection{Example 1: Different App Theme}

\begin{figure}[!h]
	\centering
	\includegraphics[width=0.85\linewidth]{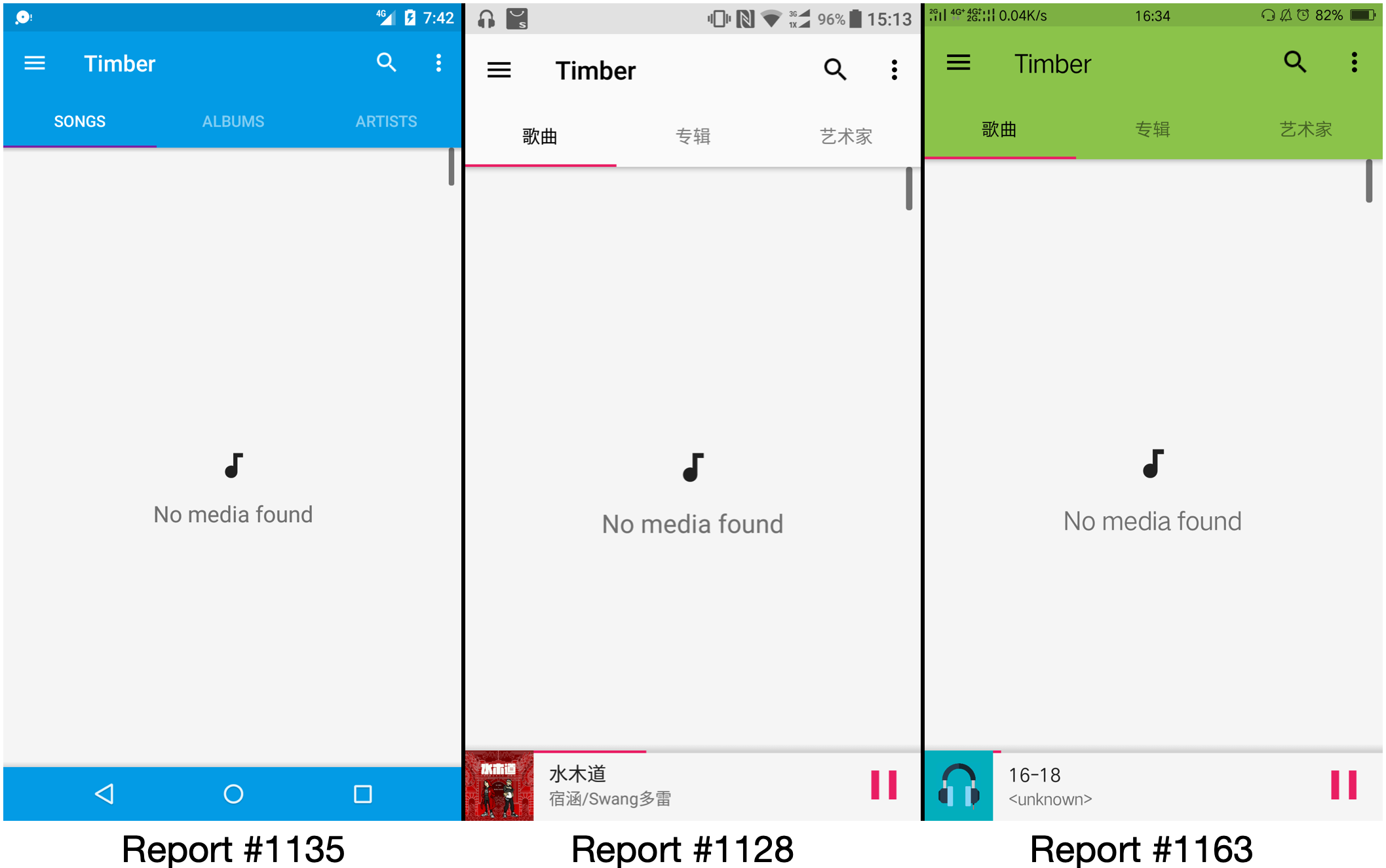}
	\caption{Example 1: Different App Theme}
	\label{fig:example1}
\end{figure}

Nowadays, apps support different themes, making it possible for users to customize the app appearance according to their preferences (\figref{fig:example1}). Moreover, the supported ``dark mode'' makes the color scheme more complex. Image feature extraction algorithms can hardly handle such complexity and will make mistakes. From the examples, we can find that the app screenshots in three reports are of blue, white, and green themes. All these three reports are reporting the loading failure of the music resource files. However, according to \cite{wang2019images}, the image color feature is one vital component of the report surrogate. App screenshots with different colors will be recognized as different screenshots.

\begin{center}\setlength{\fboxrule}{0.5pt}\fbox{\parbox{0.9\linewidth}{
	\emph{Report \#1135}: Choose a directory containing music files and ``mark as music dir'', but music files do not show when returning to main page.
	\\
	\emph{Report \#1128}: The song list cannot show the music files.
	\\
	\emph{Report \#1163}: The page shows ``No media found'' after choosing the music library.
}}\end{center}

\subsection{Example 2: Different Bugs on the Same Screenshots}

\begin{figure}[!h]
	\centering
	\includegraphics[width=0.6\linewidth]{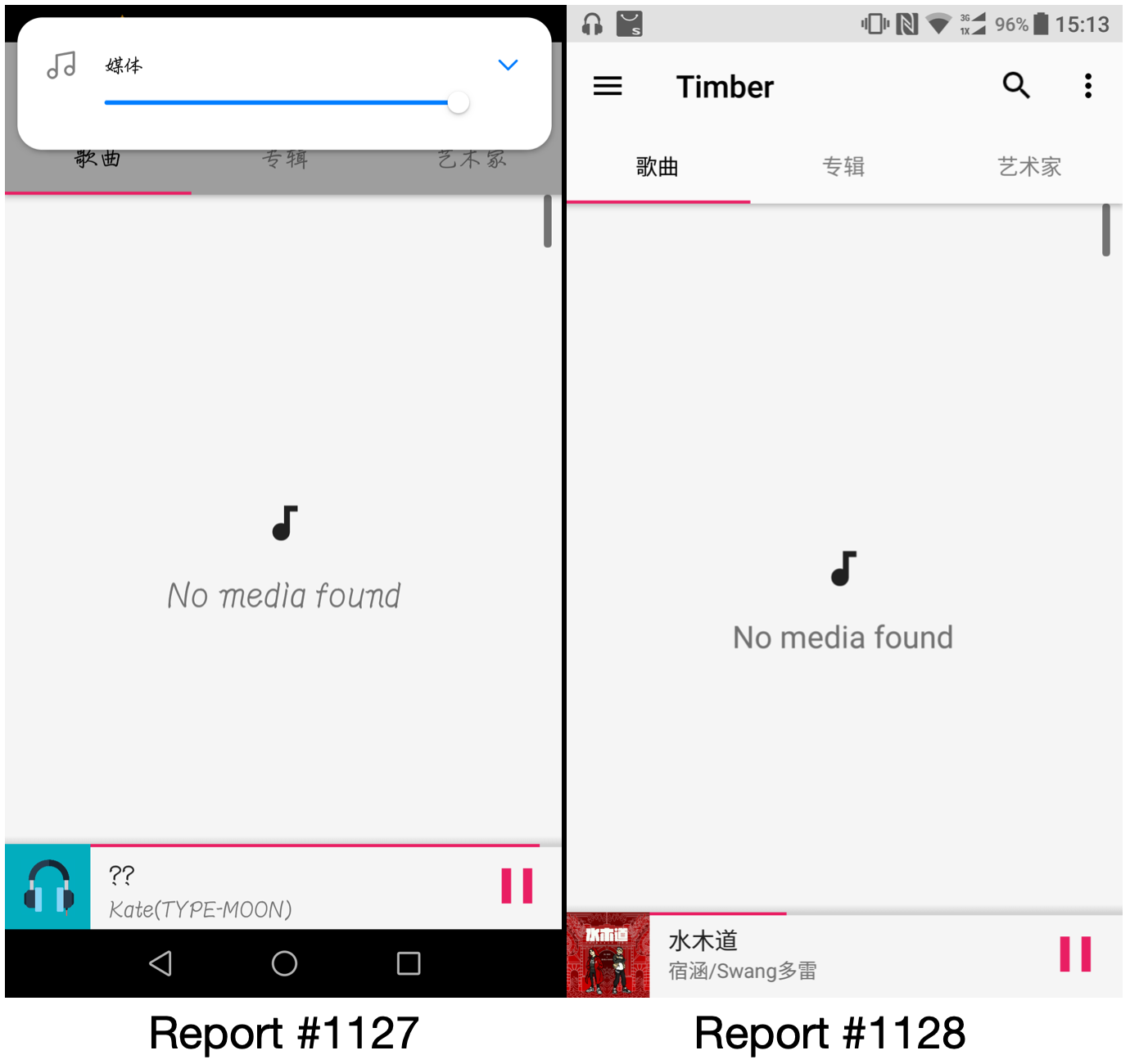}
	\caption{Example 2: Different Bugs on the Same Screenshots}
	\label{fig:example2}
	\vspace{-0.2cm}
\end{figure}

As shown in \figref{fig:example2}, the two reports use the screenshots of the same app activity, and the image feature extraction algorithm will assign a high similarity between these two screenshots. However, according to the bug description, the two reports are describing completely different bugs. In \approach, for \emph{Report \#1128}, we can extract the text ``no media found''; for \emph{Report \#1127}, we can extract the volume widget besides the prompt information, and \approach can identify the different problems.

\begin{center}\setlength{\fboxrule}{0.5pt}\fbox{\parbox{0.9\linewidth}{
	\emph{Report \#1127}: When the headphones are plugged in, the volume is automatically increased while playing.
	\\
	\emph{Report \#1128}: The song list cannot show the music files.
}}\end{center}

\subsection{Example 3: Same Bug on Different Screenshots}
\label{sec:example3}

\begin{figure}[!h]
	\centering
	\includegraphics[width=0.6\linewidth]{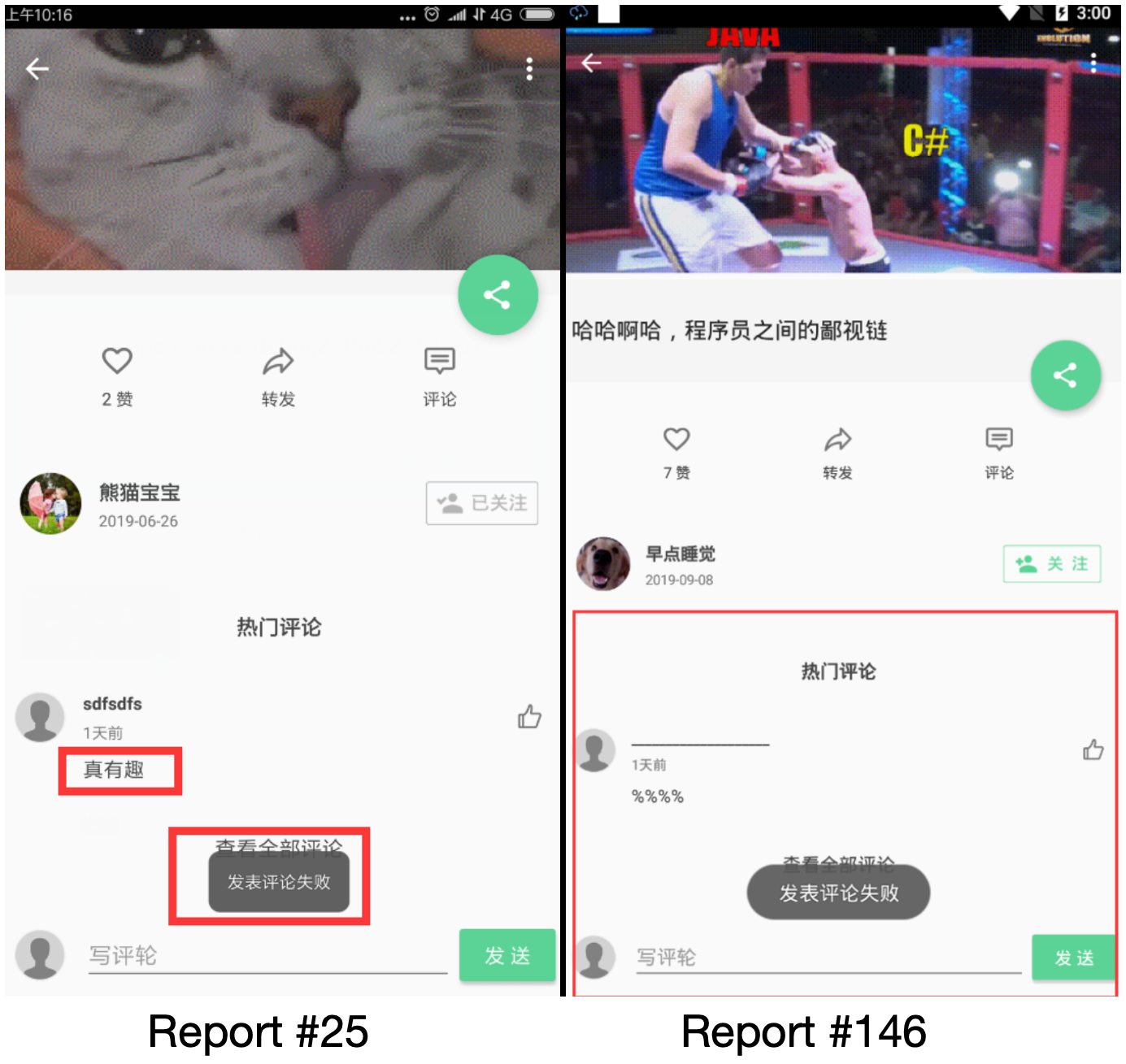}
	\caption{Example 3: Same Bug on Different Screenshots}
	\label{fig:example3}
	\vspace{-0.2cm}
\end{figure}

As shown in \figref{fig:example3}, the \texttt{ImageView} widget on the top is of different contents, and it occupies a large proportion of the entire page. Also, the comments are different due to different testing time. Therefore, existing approaches will consider the two screenshots are of low similarity which will pull down the whole similarity even if the textual descriptions are with high similarity. With \approach, we can extract the pop-up information on the bottom, saying ``comment failed'', and assign a high similarity to the two reports. Such pop-ups are considered as a quite significant widget that contains the bug.

\begin{center}\setlength{\fboxrule}{0.5pt}\fbox{\parbox{0.9\linewidth}{
	\emph{Report \#25}: The page reminds failure after submitting comment, but the submission shows in the list.
	\\
	\emph{Report \#146}: When inputing the comment, the page reminds submit failure, but when reentering the page, the comment has been in the list.
}}\end{center}

\section{Approach}
\label{sec:method}

\begin{figure*}[!h]
	\centering
	\includegraphics[width=0.9\linewidth]{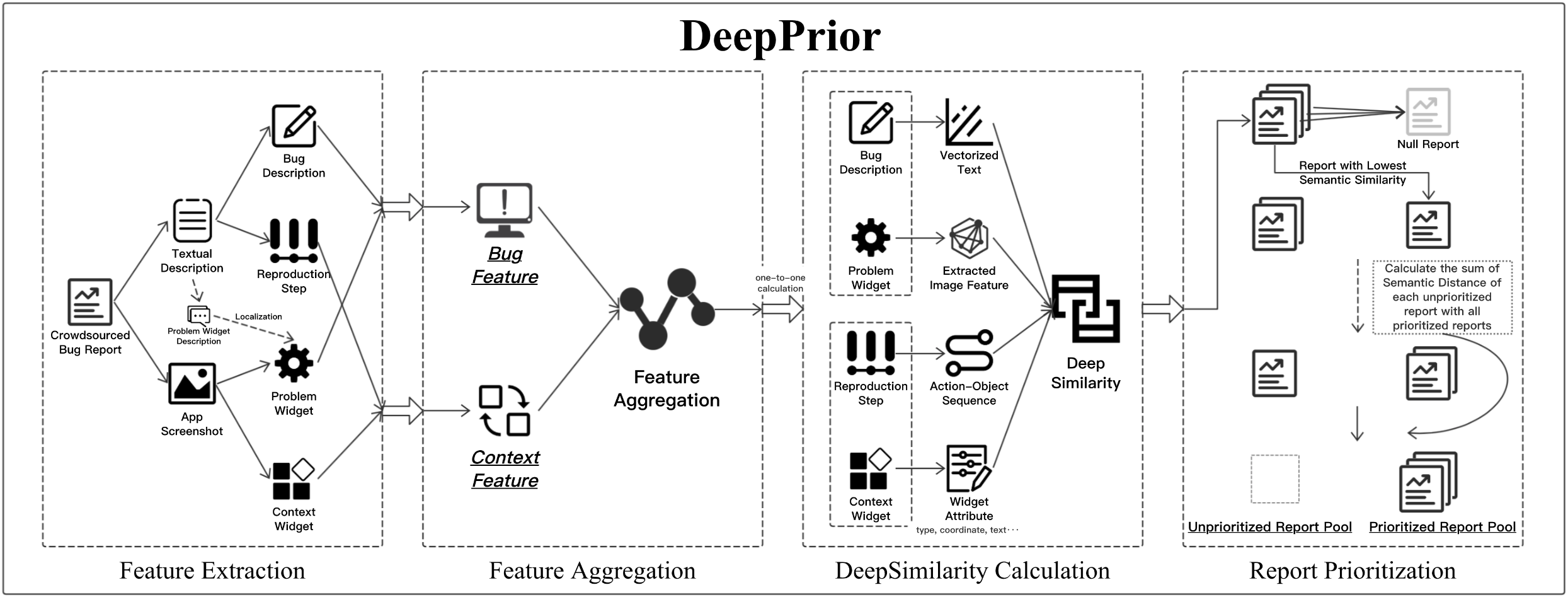}
	\caption{\approach Framework}
	\label{fig:framework}
\vspace{-0.3cm}
\end{figure*}

This section presents the details of \approach, which means prioritizing crowdsourced test reports via deep screenshot understanding. \approach consists of 4 stages, including feature extraction, feature aggregation, \similarity calculation, and report prioritization. In the first stage, we collect 4 different types of report features from both \textbf{app screenshots} and \textbf{textual descriptions}. We then aggregate the extracted features into a \deepft, which includes \bft and \cft. Based on the \deepft, we design an algorithm to calculate the \similarity between every two test reports. Based on the pre-defined rules (details in \secref{sec:prioritization}), we prioritize the test reports according to the \similarity. The general framework of the \approach approach can be referred to in \figref{fig:framework}.

\subsection{Feature Extraction}

The first and most important step is feature extraction. In this step, we analyze the app screenshots and textual descriptions in the crowdsourced test reports separately.

\subsubsection{\textbf{Features from App Screenshot}}
\label{sec:image}

App screenshots are vital in crowdsourced test reports. Crowdworkers are required to take screenshots while the bugs occur to better illustrate the bug. As described in \cite{wang2019images}, texts can be confusing, because textual descriptions can only provide limited information. Therefore, screenshots are taken into consideration to provide much more information besides textual descriptions. In a screenshot, there exist many different widgets, and some widgets can prompt the bug information. Therefore, the deep understanding of the screenshots mainly rely on the widgets. In \approach, we use both CV technologies and deep learning (DL) technologies to extract all the widgets and analyze their information. DL technologies are powerful and CV technologies can process tasks in a larger variety \cite{o2019deep}.

\textbf{\textit{Problem Widget.}}
An app activity\footnote{https://developer.android.com/reference/android/app/Activity} can be seen as an organized widget\footnote{https://developer.android.com/reference/android/widget/package-summary} set. Generally speaking, in crowdsourced testing tasks, the bugs that can be found by crowdworkers are sure to be revealed through the widgets. Therefore, it is important to locate the widget that triggers the bug and distinguish this widget, which is what we define as Problem Widget ($W_P$), from other widgets. In order to distinguish the Problem Widget, we analyze the textual descriptions. In crowdsourced test reports, crowdworkers will point out which widget is operated before the bug occurs. As shown in \secref{sec:text}, we can extract the problem widget from the textual descriptions, and to locate the Problem Widget, we adopt two different strategies for different situations:

\begin{itemize}
	\item If the extracted widget contains texts, we will match the texts from the widget screenshot and the textual description. The matched widget will be considered as the Problem Widget.
	\item If there are no texts on the widgets or the text matching fails, we will feed the extracted widgets into a deep neural network to identify the simple widget intention. The deep neural network is modified from the research of Xiao \etal \cite{xiao2019iconintent}. The model encodes the widget screenshot into a feature vector with a convolutional neural network (CNN). The output is a short text fragment decoded from the feature vector with a recurrent neural network (RNN), and the text fragment describes the widget intent.
\end{itemize}

\textbf{\textit{Context Widget.}}
Besides the Problem Widget, the widget set representing the app screenshot also contains many more widgets that make up the context, which is also critical to deep image understanding. From the early stage survey, we find that the situations are common when app activities are entirely different even if the problem widget, the reproduction step (the activity launching path), and the bug description are the same (like motivating example in \secref{sec:example3}). At this time, the context widget is significant to identify the differences. Therefore, we collect the rest of the widgets as Context Widget ($W_C$). For each context widget, we feed the widget screenshot into a convolutional neural network to identify its type. The amount for each type consists of a 14-dimensional vector.

The convolutional neural network is capable of identifying 14 different types of most widely used widgets,
including \texttt{Button} (BTN), \texttt{CheckBox} (CHB), \texttt{CheckTextView} (CTV), \texttt{EditText} (EDT), \texttt{ImageButton} (IMB), \texttt{ImageView} (IMV), \texttt{ProgressBarHorizontal} (PBH), \texttt{ProgressBarVertical} (PBV), \texttt{RadioButton} (RBU), \texttt{RatingBar} (RBA), \texttt{SeekBar} (SKB), \texttt{Switch} (SWC), \texttt{Spinner} (SPN), \texttt{TextView} (TXV)\footnote{the short name is used in this paper for convenience}. To train the neural network, we collect 36,573 widget screenshots that evenly distribute in 14 types. The ratio of the training set, validation set, and test set is 7:1:2, which is a common practice for an image classification task. The neural network is composed of multiple \texttt{Convolutional} layers, \texttt{MaxPooling} layers, and \texttt{FullyConnected} layers. \texttt{AdaDelta} algorithm is used as the optimizer, and this model adopts the \texttt{categorical\_crossentropy} loss function. 

\subsubsection{\textbf{Features from Textual Description}}
\label{sec:text}

Besides app screenshots, textual descriptions can provide bug information more intuitively and directly. Also, textual descriptions can make a positive supplement for app screenshots. In \approach, we adopt NLP technologies, specifically DL algorithms, to process the textual descriptions in the test reports.

In the textual description, crowdworkers are required to describe the bug in the screenshot and provide the reproduction step, which is the operation sequence from the app launching to the bug occurrence. However, on most crowdsourced testing platforms, the bug descriptions and the reproduction steps are mixed together, and crowdworkers are not required to obey specific patterns due to the great diversity of their professional capability \cite{yu2019crowdsourced}. Therefore, it is complex to distinguish bug descriptions from reproduction steps. In order to handle this problem, we adopt the TextCNN model \cite{kim2014convolutional}. 

The TextCNN model can complete sentence-level classification tasks with pre-trained word vectors. Before feeding the texts into the model, we pre-process the data. The textual descriptions of the test reports are segmented into sentences, and then we use \texttt{jieba} library\footnote{https://github.com/fxsjy/jieba} to segment sentences into words, and then we filter out the stop words according to a stop word list\footnote{https://github.com/goto456/stopwords}. After the pre-processing, we feed the texts into a \texttt{WordEmbedding} layer. In this layer, the texts are transformed into 128-dimensional vectors using a Word2Vec model \cite{mikolov2013distributed}. Afterwards, we adopt several \texttt{Convolutional} layers and \texttt{MaxPooling} layers to extract the text features. In the last layer, we use \texttt{SoftMax} activation function and get the probability of whether each sentence is a bug description or a reproduction step. Finally, we merge all the sentences classified as bug description or reproduction step. To train the TextCNN model, we form a large-scale text classification dataset composed of 2,252 bug descriptions and 2,088 reproduction steps. We set the ratio of the training set, validation set, and test set as 6:2:2, following the common practice.

\textbf{\textit{Bug Description.}}
Bug descriptions are always in the form of a short sentence. Therefore, we represent the sentence with a vector, which is also encoded using the Word2Vec model. Most of the bug descriptions are following some specific kind of patterns, like ``\textit{apply \underline{SOME} operation on \underline{SOME} widget, and \underline{SOME} unexpected behavior happens}'', so even if the specific words would vary, it is effective to extract such feature. 

One more important process is to extract the description of the problem widget in order to help localize the problem widget. To achieve this goal, we use text segmentation algorithms based on HMM (Hidden Markov Model) models \cite{rabiner1989tutorial} and analyze the part-of-speech of each part of the bug descriptions after text segmentation. Then, we extract the object components as the basis for problem widget localization, and such object components of the sentences are the widgets that trigger the bugs. After acquiring the objects, we use the strategies introduced above to localize the problem widget.

\textbf{\textit{Reproduction Step.}}
In addition to the bug description, another significant part of the textual description is the reproduction step. The reproduction step is a series of operations, describing the user's operations from the app launching to the bug occurrence. For sentences classified to the reproduction step class, we process in the initial order in the reports. We use the same NLP algorithms to make text segmentation and analyze the part-of-speech for each text segment for each sentence. Then, the action part and the object part are collected to form the ``action-object'' pair. Then, we concatenate the ``action-object'' pairs to an ``action-object'' sequence. Also, besides the action words and the objects, we add some complementary information for some specific operations. For example, suppose one operation is a typing action, we will add the input content as the supplementary information, because different test inputs can lead to different consequences and make the app directed to different activities. Finally, after the formalizing processing, we can obtain the Reproduction Step from the textual descriptions.

\subsection{Feature Aggregation}

After acquiring all the features both from app screenshots and textual descriptions, we aggregate them into two feature categories: \bft (BFT) and \cft (CFT). \bft refers to the features that directly reflect or describe the bug in the crowdsourced test report, while \cft is assembled by the features that provide a depiction for the environment of the bug occurrence.

\subsubsection{\bft (BFT)}

\bft can directly provide information about the bugs. Since a crowdsourced test report is composed of the app screenshot and the textual description, both components contain critical information about the occurring bug. In the app screenshot, we extract the Problem Widget, which is a widget screenshot. \approach can extract such information automatically. In the textual description, the Bug Description part directly describes the bug. Therefore, with a balanced consideration of the app screenshot and the textual description, we aggregate the Problem Widget and the Bug Description as the \bft.

\subsubsection{\cft (CFT)}

\cft includes the features that construct a thorough context for the bug occurrence. In the app screenshot, the Context Widget consists of all the widgets expect the Problem Widget. In the textual description, the Reproduction Step information is taken into consideration because it provides the full operation path from the app launching to the bug occurrence, and it can help identify whether the bugs of two test reports are on the same app activity. Therefore, Context Widget and Reproduction Step are aggregated together to form the \cft.

\subsubsection{Feature Aggregation}

With \bft and \cft, we can aggregate all the obtained features from both app screenshots and textual descriptions of the crowdsourced test reports into the final \deepft. We have a deep screenshot understanding for app screenshots instead of directly transforming the app screenshots into simple feature vectors. We also have a tighter combination between app screenshots and textual descriptions. Moreover, we take the app screenshots and textual descriptions as a whole and divide them according to their roles in bug reflection. \bft is undoubtedly important, and we hold that \cft also plays a crucial role in crowdsourced test report prioritization because the calculation of the bug similarity relies heavily on the whole context.

\subsection{\similarity Calculation}

To prioritize the crowdsourced test reports, one significant step is to calculate the similarity among all the reports. Because we are the first to introduce the deep screenshot understanding into report prioritization, we name the similarity as \similarity. As the common practice to merge different features in previous studies \cite{wang2019images}\cite{feng2016multi}, we calculate the \similarity of different features separately, and allocate different weights for the results of different features. The formal expression is as follows ($Sim$ is short for similarity):

\begin{subequations}
	\begin{equation}
		\similarity = \gamma * Sim_{\sc BFT} + (1 - \gamma) * Sim_{\sc CFT}
	\end{equation}
	\vspace{-0.1cm}
	\begin{equation}
		Sim_{\sc BFT} = \alpha * Sim_{W_P} + (1 - \alpha) * Sim_{P}
	\end{equation}
	\vspace{-0.1cm}
	\begin{equation}
		Sim_{\sc CFT} = \beta * Sim_{W_C} + (1 - \beta) * Sim_{R}
	\end{equation}
\end{subequations}

\subsubsection{\bft}

We calculate the \similarity of Problem Widget and the Bug Description separately and merge them with the weight parameter $\alpha$.

\textbf{\textit{Problem Widget.}}
Problem Widget is a widget screenshot extracted from the app screenshot according to the strategies introduced in \secref{sec:image}. To calculate the \similarity of the problem widget, we extract the image features of the widget screenshots. To extract the image features, we adopt the state-of-the-art SIFT (Scale-Invariant Feature Transform) algorithm \cite{lowe1999object}. Therefore, each widget is represented by a feature point set. SIFT algorithm has the advantage of being able to process the images with different sizes, positions, and rotation angles, which is a common phenomenon in such an era when mobile devices are of hundreds of thousands of different models. To compare and match the problem widgets from different crowdsourced test reports, we use the FLANN Library\footnote{https://github.com/mariusmuja/flann} \cite{muja2009fast}. After the calculation, we can get a score ranging from 0 to 1, and 0 means completely different, and 1 means completely the same. This score can be viewed as the \similarity of Problem Widget.

\textbf{\textit{Bug Description.}}
Bug Description is a shot sentence briefly describing the bug in the crowdsourced test report. Therefore, we use NLP technologies to encode bug descriptions. Following the approaches in previous studies, we use the Word2Vec model as the encoder. To improve the performance of the Word2Vec Model, we construct a \textit{test report keyword database}. The test report keyword database contains 8,647 keywords related to software testing, mobile app, and test report, including labeled synonyms, antonyms, and polysemies. The encoded bug description is a 100-dimensional vector. Afterward, also referring to the previous studies like \cite{wang2019images}\cite{feng2016multi}, we adopt the widely used \textit{Euclidean Metric} algorithm to calculate the \similarity of bug descriptions of different test reports in pair. To unify values of different scales, we normalize each result $x$ to [0, 1] interval with the function $\frac{x-min}{max-min}$, where $max$ is the maximum value of all results and $min$ is the minimum value of all results.

\subsubsection{\cft}

We also calculate the \similarity of Context Widget and the Reproduction Step separately and merge them with a weight parameter $\beta$.

\textbf{\textit{Context Widget.}}
Context Widget is also a very important component of the whole context of the occurring bug. To have a deep understanding of the app screenshots, specifically the widgets on the app screenshot, we use a convolutional neural network to identify the widget type for each extracted widget screenshot and form a vector containing the number of the 14 types of the widgets. Afterward, we use the \textit{Euclidean Metric} algorithm to calculate the distance of the acquired 14-dimensional vectors. We consider the absolute amount of the widgets for each type and all the widgets' distribution. The result of the \textit{Euclidean Metric} algorithm (from 0 to 1) is considered as the Context Widget \similarity. 

\textbf{\textit{Reproduction Step.}}
Reproduction Step is transformed into an ``action-object'' sequence during the feature extraction. To calculate the \similarity of the ``action-object'' sequence, we adopt the Dynamic Time Warping (DTW) \cite{silva2016speeding} algorithm to process the to-compare ``action-object'' sequences. DWT algorithm is most widely known for the capability of automatic speech recognition. In this paper, we adapt the DWT algorithm to process the operation path that triggers the bug in the corresponding crowdsourced test reports. DWT algorithm can measure the similarity of the temporal sequences, especially the temporal sequences that may vary in ``speed''. Specifically speaking, the ``speed'' in our task refers to the situation that the different user operations can reach the same app activity through a different path. Compared with other track similarity algorithms, DTW has a better matching effect because it can process the sequences with different lengths, which is suitable for processing the ``action-object'' sequences.

\subsection{Report Prioritization}
\label{sec:prioritization}

After aggregating the \deepft, and defining the \similarity calculation rule, we start to prioritize the crowdsourced test reports. First, we construct two null report pools: the unprioritized report pool and the prioritized report pool. All the crowdsourced test reports are put into the unprioritized report pool initially.

Different from the strategy adopted in \cite{feng2016multi}, where a report is randomly chosen as the initial report, we think all reports should be treated equally, and the randomly chosen report is likely to affect the final prioritization. Therefore, to formalize and unify the prioritization algorithm, we introduce the concept of \textit{NULL Report}, which also contains four features. 

\begin{algorithm}[h]
	\caption{Crowdsourced Test Report Prioritization}
	\label{alg:prioritization}
	\begin{algorithmic}[1]
		\Require Crowdsourced Test Report Set $ R_{initial} $
		\Ensure Prioritized Crowdsourced Test Report Set $ P $ 
		\State initiate unprioritized report pool $ U \leftarrow R_{initial} $ 
		\State initiate prioritized report pool $ P = \emptyset $
		\State initiate target report $ r_t $
		\State initiate NULL Report $ r_{null} $ 
		\State $P$.append($ r_{null} $)
		\While{ $ |U| \neq 0 $ }
			\State initiate $similarity = 2$
			\For{each $r \in U$}
				\For{each $r_p \in P$}
					\State initiate $similarity_r$ = 2
					\State $Sim_{BFT} = \alpha * Sim_{W_P} + (1 - \alpha) * Sim_{P}$
					\State $Sim_{CFT} = \beta * Sim_{W_C} + (1 - \beta) * Sim_{R}$
					\State calSemSim()$=\gamma * Sim_{\sc BFT} + (1-\gamma) * Sim_{\sc CFT}$
					\State $similarity_r ~+=$ calSemSim($r$, $ r_p $)
					\If{calSemSim($r$, $ r_p $) $ < similarity_r$ }
						\State $similarity_r$ = calSemSim($r$, $ r_p $)
					\EndIf
				\EndFor
				\If{$similarity_r < similarity$}
					\State $ r_t = r$
					\State $similarity = similarity_r$
				\EndIf  
			\EndFor 
			\State $P$.append($ r_t $)
			\State $U$.remove($ r_t $)
		\EndWhile
		\State \Return $ P $
	\end{algorithmic}
\end{algorithm}

\begin{itemize}
	\item \underline{Problem Widget}: the screenshot of the problem widget is essentially a 3-dimensional matrix representing the width, the height, and three color channels. Therefore, we construct the problem widget as a zero matrix. The width and the height of the zero matrix are set as the average size of all the actual crowdsourced test reports. Intuitively speaking, it is an all-black image.
	\item \underline{Bug Description}: the bug description of the \textit{NULL Report} is directly set as an empty string, and since the string length is 0, obviously it contains no words, after the Word2Vec processing, the feature vector will be a 100-dimensional vector of all `0's.
	\item \underline{Context Widget}: for the context widget of the \textit{NULL Report}, we directly construct the vector representing the numbers of the 14 different types of widgets, and all elements are 0. This represents that there are ``no'' widgets on the app screenshot of the crowdsourced test report.
	\item \underline{Reproduction Step}: the reproduction step of the \textit{NULL Report} is also set as an empty string, and the ``action-object'' sequence is also with a length of 0.
\end{itemize}

The primary consensus for prioritization is to reveal all the bugs as early as possible under the circumstances when some reports would describe the problems repetitiously \cite{feng2016multi}\cite{chen2010adaptive}\cite{jiang2009adaptive}. 

Therefore, it is important to provide as many reports describing different bugs as possible for the developers early. Based on this idea, we design our prioritization strategy as follows, and the formal expression is presented in \algopref{alg:prioritization}.

First, we construct the \textit{NULL Report} according to the rules. mentioned above and append the \textit{NULL Report} to the null prioritized report pool. Following is a iterative process. We calculate the \similarity of each unprioritized report to the whole prioritized report pool, which is defined as the minimum \similarity of the unprioritized report to all the reports in the prioritized report pool. The report with the lowest \similarity with the prioritized report pool will be moved to the prioritized report pool.

\section{Evaluation}
\label{sec:eval}

\subsection{Experimental Setting}

To evaluate the proposed \approach, we design an empirical experiment. To complete the experiment, we collect 536 crowdsourced test reports from 10 different mobile apps (details in \tabref{tbl:app}). The apps are labeled from A1 to A10, and the number of test reports of different apps ranges from 10 to 152. We also invite software testing experts to manually classify the test reports according to the bugs they describe, and the average number of reports in a bug category is 8.06.

\begin{table}[!h]
\centering
\caption{Experiment App}
	\begin{tabular}{c|c|c|c|c}
		App No. & App Category & Report \# & \tabincell{c}{Bug \\ Category \#} &
		\tabincell{c}{Report per \\ Category} \\ \hline \hline
		A1  & Finance       & 134 & 9  & 14.89 \\
		A2  & System        & 29  & 6  & 4.83  \\
		A3  & Reading       & 23  & 4  & 5.75  \\
		A4  & Reading       & 152 & 8  & 19.00 \\
		A5  & System        & 75  & 7  & 10.71 \\
		A6  & Finance       & 26  & 5  & 5.20  \\
		A7  & Finance       & 10  & 2  & 5.00  \\
		A8  & Music         & 51  & 8  & 6.38  \\
		A9  & Life          & 12  & 3  & 4.00  \\
		A10 & System        & 24  & 5  & 4.80  \\ \hline \hline
		\multicolumn{2}{c|}{Total} & 536 & Average & 8.06
	\end{tabular}
\label{tbl:app}
\vspace{-0.5cm}
\end{table}

Based on the crowdsourced test report dataset, we build three specific datasets to better support the evaluation, including \textbf{1)} a large-scale widget image dataset, \textbf{2)} a large-scale test report keyword set, and \textbf{3)} a large-scale text classification dataset. All the 4 dataset build up the integrated dataset group.

In total, we design three research questions (RQ) to evaluate the proposed test report prioritization approach, \approach. 

\begin{itemize}
	\item \textbf{RQ1}: How effective can \approach identify the widget type extracted from the app screenshots?
	\item \textbf{RQ2}: How effective can \approach classify the textual descriptions from the crowdsourced test reports?
	\item \textbf{RQ3}: How effective can \approach prioritize the crowdsourced test reports?
\end{itemize}

\subsection{RQ1: Widget Type Classification}

The first research question is set to evaluate the effectiveness of our processing to the app screenshots. The most important component of app screenshot processing is the widget extraction and classification. Therefore, we evaluate the accuracy of the widget type classification CNN. 36,573 different widget images are collected from real-world apps, and the images have an even distribution in 14 categories. 

The details of the CNN is presented in \secref{sec:image}. The dataset is divided by the ratio 7:2:1 into the training set, the validation set and the test set, as the common practice. After the CNN model is trained, we evaluate the accuracy on the test set. The overall accuracy of widget type classification reaches 89.98\%. Specifically speaking, we use \textit{precision}, \textit{recall} and \textit{F-Measure} values to evaluate the network. The calculating formula can be seen as follows, where \textit{TP} means true positive samples, \textit{FP} means false-positive samples, \textit{TN} means true negative samples, \textit{FN} means false negative samples.

\begin{equation}
\vspace{-0.2cm}
	precision = \frac{TP}{TP + FP},~~~~recall = \frac{TP}{TP + FN}
\end{equation}

\begin{equation}
	\frac{2}{F-Measure} = \frac{1}{precision} + \frac{1}{recall}
\end{equation}

The evaluation results can be seen in \tabref{tbl:widgettype}. The \textit{precision} value reaches an average of 90.05\%, and the lowest \textit{precision} is 74.36\% and the highest is 99.81\%. For \textit{recall}, which measures the total amount of relevant instances that are actually retrieved, the average values is 89.98\%, and the \textit{recall} values range from 70.83\% to 100.00\%. \textit{F-Measure} is a harmonic mean of the \textit{precision} and \textit{recall}, and it reaches an average of 89.92\%. The above results reflect the outstanding capability of the proposed classifier.

\begin{table}[!h]
	\centering
	\caption{Widget Type Classification}
	\begin{tabular}{c|c|c|c}
		Widget Type & precision & recall   & F-Measure  \\ \hline \hline
		BUT         & 82.30\%   & 76.53\%  & 79.31\% \\
		CHB         & 96.35\%   & 94.89\%  & 95.61\% \\
		CTV         & 93.27\%   & 92.73\%  & 93.00\% \\
		EDT         & 74.66\%   & 87.81\%  & 80.70\% \\
		IMB         & 76.73\%   & 85.26\%  & 80.77\% \\
		IMV         & 74.36\%   & 72.01\%  & 73.17\% \\
		PBH         & 98.65\%   & 93.78\%  & 96.15\% \\
		PBV         & 94.35\%   & 99.81\%  & 97.00\% \\
		RBU         & 94.17\%   & 93.45\%  & 93.81\% \\
		RBA         & 99.05\%   & 99.81\%  & 99.43\% \\
		SWC         & 98.85\%   & 97.73\%  & 98.29\% \\
		SKB         & 99.23\%   & 94.89\%  & 97.01\% \\
		SPN         & 99.81\%   & 100.00\% & 99.90\% \\
		TXV         & 78.95\%   & 70.83\%  & 74.67\% \\ \hline \hline
		Average     & 90.05\%   & 89.97\%  & 89.92\% \\ 
	\end{tabular}
	\label{tbl:widgettype}
\vspace{-0.3cm}
\end{table}

We also have an in-depth insight into the results. We find that there are two groups of widgets that are easy to be confused. The first group includes \texttt{ImageButton} and \texttt{ImageView}. It is easy to understand that from a visual perspective, these two types can hardly be identified. The only difference between these 2 types is that \texttt{ImageButton} can trigger an action while \texttt{ImageView} is a simple image. However, one important thing to mention is that the in app design, developers can add a hyperlink to the \texttt{ImageView} widget to realize the equivalent effect. The second group includes \texttt{Button}, \texttt{EditText} and \texttt{TextView}. These three widgets are all a fixed area containing a text fragment, which is also visually similar and hard to identify even for humans. Moreover, some special renderings make the widgets even harder to identify. According to our survey, we find that these two confusing groups will not affect much, and the widgets can be treated as equivalent from both visual perspective and function perspective.

\textbf{\textit{Results for RQ1}}: The overall accuracy of the CNN to classify the widget types reaches 89.98\%. For each specific type, the average \textit{precision} is 90.05\%, the \textit{precision} is 74.36\% and the \textit{F-Measure} is of 89.92\%. Also, according to our survey on real test reports, even if some types that are with low precision, their visual and functional features will not negatively affect \approach.

\subsection{RQ2: Textual Description Classification}

In the processing of the textual description, we classify into two categories: bug description and reproduction step. Different textual descriptions are considered as different report features. To classify the textual descriptions, we segment the textual descriptions into sentences. Then, we feed the sentences into a TextCNN model to complete the task, and the details are presented in \secref{sec:text}. Also, to better train and evaluate the network, we build a large-scale text classification dataset. The dataset contains 4,340 pieces of labeled textual segments, including 2,252 pieces of bug descriptions and 2,088 pieces of reproduction steps. The dataset is divided into training set, validation set and test set at the ratio of 7:2:1.

\begin{table}[!h]
	\centering
	\caption{Text Classification }
	\begin{tabular}{c|c|c|c}
	                   & precision & recall  & F-Measure \\ \hline \hline
	Bug Description    & 98.46\%   & 97.81\% & 98.13\%   \\
	Reproduction Step  & 97.95\%   & 98.53\% & 98.24\%   \\ \hline \hline
	Average            & 98.21\%   & 98.17\% & 98.19\%   \\ 
	\end{tabular}
	\label{tbl:textclassification}
\vspace{-0.3cm}
\end{table}

The results on the test set can be seen from \tabref{tbl:textclassification}, and the overall accuracy of the model reaches 96.65\%. More specifically, we use the same measurements like the evaluation in RQ1, including \textit{precision}, \textit{recall} and \textit{F-Measure}. The average \textit{precision} of 2 types of textual description reaches 98.21\%, the average \textit{recall} reaches 98.17\%, and the he average \textit{F-Measure} reaches 98.19\%. The result is quite promising, and we also manually check the textual descriptions. We find that compared with reproduction steps, bug descriptions tend to contain bug-related words, such as ``crash'', ``flashback'', ``missing element'', ``wrong'', ``fail'', ``no response'', \etc~While the reproduction steps contain just the operations, target widgets and the corresponding responses.

\textbf{\textit{Results for RQ2}}: The overall accuracy of text classification reaches 96.65\%, and the \textit{precision}, \textit{recall} and \textit{F-Measure} are all over 98\%. Such results show \approach's excellent capability to analyze textual descriptions, which also lays a solid foundation for the crowdsourced test report prioritization. 

\begin{table*}[]
\centering
\caption{\approach Report Prioritization Result and Comparison}
\begin{tabular}{c|c||c||c|c|c||c|c||c|c}
App No. & {\sc \textbf{IDEAL}} & {\sc \textbf{DeepPrior}} 
& {\sc \textbf{BDDiv}} & \tabincell{c}{{\sc \textbf{DeepPrior}} \\ vs. {\sc \textbf{BDDiv}}}
& \tabincell{c}{Overhead \\ Comparison} 
& {\sc \textbf{Image}} & \tabincell{c}{{\sc \textbf{DeepPrior}} \\ vs. {\sc \textbf{Image}}}
& {\sc \textbf{Random}} & \tabincell{c}{{\sc \textbf{DeepPrior}} \\ vs. {\sc \textbf{Random}}} \\ \hline \hline
A1      & 0.974 & 0.927 & 0.914 & 1.36\%  & 68.75\% & 0.926 & 0.09\%  & 0.805 & 15.15\% \\
A2      & 0.931 & 0.839 & 0.822 & 2.10\%  & 33.40\% & 0.805 & 4.29\%  & 0.655 & 28.07\% \\
A3      & 0.957 & 0.865 & 0.827 & 4.65\%  & 35.49\% & 0.721 & 20.00\% & 0.692 & 25.00\% \\
A4      & 0.980 & 0.933 & 0.941 & -0.87\% & 83.29\% & 0.845 & 10.42\% & 0.794 & 17.51\% \\
A5      & 0.967 & 0.898 & 0.894 & 0.43\%  & 56.14\% & 0.892 & 0.64\%  & 0.751 & 19.52\% \\
A6      & 0.942 & 0.827 & 0.765 & 8.04\%  & 41.89\% & 0.827 & 0.00\%  & 0.619 & 33.54\% \\
A7      & 1.000 & 1.000 & 1.000 & 0.00\%  & 32.18\% & 1.000 & 0.00\%  & 0.750 & 33.33\% \\
A8      & 0.941 & 0.850 & 0.826 & 2.97\%  & 34.38\% & 0.858 & -0.86\% & 0.694 & 22.61\% \\
A9      & 0.958 & 0.931 & 0.875 & 6.35\%  & 23.97\% & 0.847 & 9.84\%  & 0.681 & 36.73\% \\
A10     & 0.938 & 0.863 & 0.763 & 13.11\% & 35.12\% & 0.854 & 0.98\%  & 0.621 & 38.93\% \\ \hline \hline
Average & 0.956 & 0.893 & 0.863 & 3.81\%  & 44.46\% & 0.857 & 4.54\%  & 0.706 & 27.04\% \\ 
\end{tabular}
\label{tbl:deepprior}
\vspace{-0.5cm}
\end{table*}

\subsection{RQ3: Crowdsourced Test Report Prioritization}

In this research question, we evaluate the test report prioritization effectiveness of \approach. The metric we use is the APFD (Average Percentage of Fault Detected) metric \cite{rothermel2001prioritizing}, which is also used by Feng \etal to prioritize crowdsourced test reports \cite{feng2016multi}. In the formula, the $T_{fi}$ means the index of the report that first finds the bug $i$, the $n$ is the total report number, and the $M$ is the total number of revealed bugs. 

\begin{equation}
	APFD = 1-\frac{\sum_{i=1}^{M}{T_{fi}}}{n \times M} + \frac{1}{2 \times n}
\end{equation} 

To better illustrate the advantage of \approachExp, we compare \approachExp with the following prioritizing strategies:

\begin{itemize}
	\item {\sc \textbf{Ideal}}: This strategy is the best prioritization on theory, which means that developers can review all the bugs revealed by the reports in the shortest time.
	\item {\sc \textbf{Image}}: This strategy uses only the results of deep image understanding of \approachExp to rank the test reports because deep image understanding is a significant part of our research.
	\item {\sc \textbf{BDDiv}}: This strategy refers to the algorithm in Feng \etal's work \cite{feng2016multi}, which is also the state-of-the-art approach for crowdsourced test report prioritization.
	\item {\sc \textbf{Random}}: The {\sc Random} strategy refers to the situation without any prioritization strategy. 
\end{itemize}

For \approachExp and {\sc \textbf{Image}} strategy, we run once because of the stability of our approach, and the trained model will not produce different results for various attempts; for {\sc \textbf{Ideal}} strategy, we manually calculate the APFD because for a fixed report cluster, it is a fixed value; for {\sc \textbf{BDDiv}} strategy, we run 30 times and calculate the average value as the original paper \cite{feng2016multi}; and for {\sc \textbf{Random}} strategy, we run 100 times to eliminate the effect of the occasional circumstances.

First, we compare \approachExp with {\sc \textbf{Random}} strategy. As shown in the  \tabref{tbl:deepprior}, we find \approachExp outperforms {\sc \textbf{Random}} strategy much, ranging from 15.15\% to 38.93\%, and the average improvement reaches 27.04\%. This shows a superiority of \approachExp.

Then we compare the results of \approachExp with the single {\sc \textbf{Image}} strategy. The average improvement of \approachExp is 4.54\%, and in 2 apps (A3 and A4), \approachExp outperforms much. For A8, \approachExp is slightly weaker than {\sc \textbf{Image}} strategy. We review the reports of A8 and find that the textual descriptions are not well written, and cannot positively help the prioritization of the reports. Generally speaking, the results prove the necessity of combining both text analysis and deep image understanding, and a single strategy will compensate each other's drawbacks and improve the prioritization accuracy.

Also, we make a comparison between \approachExp and {\sc \textbf{BDDiv}} strategy, which is the state-of-the-art approach. According to the experiment results, \approachExp outperforms {\sc \textbf{BDDiv}} with an average improvement of 3.81\%. The improvements in some apps are especially obvious. Moreover, we record the total time overhead from reading the report cluster to the output of the prioritized reports. It shows that \approachExp uses less than half of the time of {\sc \textbf{BDDiv}}, which shows great performance superiority.

Another advantage of \approachExp over {\sc \textbf{BDDiv}} is that the \approachExp can output stable results while {\sc \textbf{BDDiv}}'s results will float. According to the detailed results of {\sc \textbf{BDDiv}} strategy (in the online package), we find that {\sc \textbf{BDDiv}} is quite volatile.

The improvements over baseline strategies vary among different apps, and some reasons account for this. First, the ``report/category'' rate for each app is different, so in the limited activity set of an app, the recurring of the same activities become frequent. Second, different apps have various contents. For example, A1 is a kids' education app, and it consists of a large number of pictures, videos, variant texts. Such a situation makes it much more complex to extract useful text information and have a thorough understanding of the app screenshot. As a result, the matric would decrease.

\textbf{\textit{Results for RQ3}}: \approachExp's capability of prioritizing crowdsourced test reports is excellent, it outperforms the state-of-the-art approach, {\sc \textbf{BDDiv}}, with less than half the overhead. Also, the specific experiment of the {\sc \textbf{Image}} strategy shows the effectiveness of our deep screenshot understanding algorithm. Compared with the state-of-the-art approach, \approachExp performs much more stable.

\subsection{Threats to Validity}

\textbf{The categories of the apps in this experiment are limited}. Our 15 experimental apps cover eight different categories (according to app store taxonomy). The coverage is limited. However, we want to emphasize that due to our deep screenshot understanding involves the layout characterization to the app activity, the \approach is only suitable for analyzing the apps with a grid layout or a list layout. We also limit our claim within apps of such layouts.

\textbf{The enrollment of the crowdworkers is also uncontrolled}. The crowdworkers' capability is uncontrolled, and low-quality reports may occur. However, even if the quality of some reports is low, \approach can identify the bug it is describing if it actually contains one. If not, \approach will categorize the report as a single category, and will not affect the prioritization of other reports.

\textbf{The datasets we construct are of Chinese language}. The language of the datasets may be another threat, but NLP and OCR technologies are quite strong. If we replace the text processing engine with that of another language, the text processing will also be completed well, and will not have a negative impact on \approach. Moreover, the maturity of machine translation \cite{karimi2011machine} also makes it robust to process cross-language textual information.

\section{Related Work}
\label{sec:rw}

\subsection{Crowdsourced Testing}

Crowdsourced testing has been a mainstream testing strategy. It is significantly different from traditional testing. Testing tasks are distributed to a large group of crowdworkers of different locations and have widely varying abilities. The most notable advantage of crowdsourced testing is the capability of simulating different using conditions and the relatively low economic cost \cite{gao2019successes}\cite{mao2017survey}. However, the openness of crowdsourced testing leads to a large amount redundant reports. The key problem is to improve the developers' efficiency to review the test reports. some researches start from selecting skillful crowdworkers to complete the tasks \cite{cui2017multi}\cite{cui2017should}\cite{xie2017cocoon}. Such a strategy is effective, while it is still hard to control because even skillful crowdworkers would loaf on the task. Therefore, we think it is more important to process the test reports instead of other factors in crowdsourced testing. Liu \etal \cite{liu2018generating} and Yu \cite{yu2019crowdsourced} proposed approaches respectively to automatically generate descriptions from screenshots for test reports based on the consensus that app screenshots are easy to acquire while the textual descriptions are hard to write for all the crowdworkers. This idea inspired us to have a deep screenshot understanding of to help better prioritize the test reports.

\subsection{Crowdsourced Test Report Processing}

Many researches have been done to process crowdsourced test reports to better help developers review the reports and fix bugs. Basic strategies include report classification, duplicate detection, and report prioritization. In this section, we will present the related works of different basic strategies.

Banerjee \etal proposed FactorLCS \cite{banerjee2012automated}, utilizing the common sequence matching, and the approach is effective on open bug tracking repositories. They also proposed a method \cite{banerjee2013fusion} with a multi-label classifier to find the ``primary'' report of a cluster of reports with high similarity. Similarly, Jiang \etal proposed TERFUR \cite{jiang2009adaptive}, a tool that clusters the test reports with NLP technologies, and they also filter out the low-quality reports. Wang \etal \cite{wang2016towards} takes the features of the crowdworkers into consideration as a feature of the test reports, and then make the clustering. Wang \etal propose the LOAF \cite{wang2016local}, which is the first considering the operation steps and result descriptions separately for report feature extraction.

More researches are done to detect the duplication of the test reports. Sun \etal \cite{sun2010discriminative} use information retrieval models to detect duplicate bug reports more accurately. Sureka \etal \cite{sureka2010detecting} adopted a character n-gram based model to complete the duplicate detection task. Prifti \etal \cite{prifti2011detecting} conducted a survey on large-scale open-sourced project test reports and proposed a method that can concentrate the search for duplicate reports on specific portions of the whole repository. Sun \etal proposed a retrieval function, REP \cite{sun2011towards}, to measure the similarity, and the function includes the similarity of non-textual fields like component, version, \etc Nguyen \etal introduced DBTM \cite{nguyen2012duplicate}, a tool that utilizes both IR-based features and topic-based features, and detects the duplications bug reports according to technical issues. Alipour \etal \cite{alipour2013contextual} had a more comprehensive analysis of the test report context and improved the detection accuracy. Hindle \cite{hindle2016contextual} makes improvements by combining contextual quality attributes, architecture terms, and system-development topics to improve bug deduplicate detection.

The above approaches, including the report classification and the duplicate detection, choose part of the test reports to represent all the test reports. However, we hold that all the reports contain valuable information even if duplicates exist. Moreover, After duplicates are detected, developers still need to review the reports to carry forward the bug processing. Therefore, we think report prioritization is a better choice.

There are also many researches on report prioritization. Zhou \etal introduced BugSim \cite{zhou2012learning}, considering both textual and statistical features, to rank the test reports. DRONE, proposed by Tian \etal \cite{tian2013drone}, is a machine learning-based approach to predict the priority of the test reports, considering different factors of the test reports. Feng \etal proposed a series of approaches, DivRisk \cite{feng2015test} and BDDiv \cite{feng2016multi}, to prioritize the test reports, and they first consider the screenshots of the test reports. Subsequently, Wang \etal \cite{wang2019images} work further and explore a more sound approach to prioritize test reports with much more attention paid to the screenshots.

Among all the above researches, only a few of them, like \cite{wang2019images} and  \cite{feng2016multi}, consider the app screenshots, which we think is a valuable factor for extracting the features to process the test reports. But such researches only treat the screenshots as simple matrixes instead of meaningful content.

\subsection{Deep Image Understanding}

Image understanding is a hotspot issue in the Computer Vision (CV) field. In this section, we mainly focus on the researches utilizing image understanding in software testing. 

Lowe \cite{lowe1999object} proposed the SIFT algorithm, which is used to match the feature points on the target images and calculate the similarity, making use of a kind of new image local features, which are invariant to image changing, including translation, scaling, and rotation. Optical character recognition (OCR) is a widely-used tool to recognize the texts, which is helpful to better understand the images based on the rich textual information. Nguyen \etal \cite{Nguyen2016Reverse} proposed REMAUI, making use of CV technologies to identify the widgets, texts, images, and even containers of the app screenshot. Moran \etal \cite{moran2018machine} proposed REDRAW based on REMAUI, which is more precise to identify the widgets and can automatically generate codes for the app UI. Similarly, Chen \etal \cite{chen2018ui} also proposed a tool to generate GUI skeletons from app screenshots with the combination of CV technologies and machine learning. Yu \etal \cite{yu2019lirat} proposed a tool named LIRAT to record and replay mobile app test scripts among different platforms with a thorough understanding of the app screenshot.

\section{Conclusion}
\label{sec:con}

This paper proposes a crowdsourced test report prioritization approach, \approach, via deep screenshot understanding. \approach transforms the app screenshots and textual descriptions into four different features, including problem widget, context widget, bug description, and reproduction step. Then, the features are aggregated into \deepft, including \bft and \cft according to their relativity to the bug. Afterwards, we calculate the \similarity based on the features. Finally, the reports are prioritized according to the \similarity with a preset rule. We also conducted an experiment to evaluate the proposed approach, and the results show that \approach outperforms the state-of-the-art approach with less than half the overhead.

\section*{Acknowledgement}

This work is supported partially by National Key R\&D Program of China (2018AAA0102302), National Natural Science Foundation of China (61802171, 61772014, 61690201), Fundamental Research Funds for the Central Universities (14380021), and National Undergraduate Training Program for Innovation and Entrepreneurship (202010284073Z).

\bibliographystyle{IEEEtran}
\bibliography{main}

\end{document}